\documentclass[twoside,10pt,a4paper]{newFNLstyle}

\usepackage{amsmath}
\usepackage{amssymb}
\usepackage{graphicx,epsfig}
\usepackage{xspace}
\usepackage{vmargin}
\usepackage{hyperref}
\usepackage{cite}

\def\etal{{\it et al.,}\xspace}

\def\ie{{\it i.e.}\xspace}
\def\eg{{\it e.g.}\xspace}

\def\ra                 {\ensuremath{\rightarrow}\xspace}

\def\hz                 {\text{~Hz}\xspace}

\newcommand{\randweb}{http://planck.lal.in2p3.fr/wiki/pmwiki.php/Softs/AbsRand}
\newcommand{\planckweb}{http://www.rssd.esa.int/index.php?project=PLANCK}
\newcommand{\oofweb}{http://www.nslij-genetics.org/wli/1fnoise/index.html}
\newcommand{\arxiv}{http://arxiv.org/}
\newcommand{\pinkaudio}{http://www.firstpr.com.au/dsp/pink-noise/}
\newcommand{\labview}{http://zone.ni.com/devzone/cda/epd/p/id/1023}
\newcommand {\fk} {\ensuremath{f_\text{knee}}\xspace}
\newcommand {\fmin} {\ensuremath{f_\text{min}}\xspace}
\newcommand {\fsamp} {\ensuremath{f_\text{sample}}\xspace}
\newcommand {\w} {\ensuremath{\omega}\xspace}
\newcommand {\wz} {\ensuremath{\w_0}\xspace}
\newcommand {\wk} {\ensuremath{\w_k}\xspace}
\newcommand {\oofa} {\ensuremath{1/f^\alpha}\xspace}
\newcommand {\ooff} {\ensuremath{1/f^2}\xspace}
\newcommand {\tz} {\ensuremath{\xrightarrow{z}}\xspace}
\newcommand {\zinv} {\ensuremath{z^{-1}}\xspace}
\newcommand {\nn} {\nonumber}

\providecommand{\serie}[1]{\ensuremath{\lbrace#1\rbrace}}

\begin{document}

\volnumpagesyear{0}{0}{000--000}{2006}
\dates{\today}{revised date}{accepted date}

\title{Generating long streams of \oofa noise}

\authorsone{S. Plaszczynski}
\affiliationone{Laboratoire de l'Acc\'el\'erateur Lin\'eaire,
\\{IN2P3-CNRS et Universit\'e Paris-Sud 11, Centre Scientifique
  d'Orsay, B.P. 34,}
\\{91898 ORSAY Cedex, France.}}
\mailingone{plaszczy@lal.in2p3.fr}

\maketitle


\keywords{$1/f$ noise, simulation, random walk, digital filtering, fractals}

\begin{abstract}
We review existing methods for generating long streams of \oofa noise
($0<\alpha\le 2$) focusing on the digital filtering of white noise.
We detail the formalism to conceive an efficient \oofa random number
generator (white outside some bounds) in order to generate
very long streams of noise without an exhaustive computer
memory load.
For $\alpha=2$ it is shown why the process is equivalent to
a random-walk and can be obtained simply
by a first order filtering of white noise.
As soon as $\alpha<2$ the problem becomes non linear 
and we show why the exact digital filtering method becomes
inefficient. Instead,
we work out the formalism of using several \ooff filters spaced
logarithmically, to approximate the spectrum at the percent level.
Finally, from work on logistic maps, we give hints on how to design generators with $\alpha>2$.
The corresponding software is available from \href{\randweb}{\texttt{\randweb}}.
\end{abstract}

\section*{Introduction}
Many physical systems exhibits some \oofa noise, \ie a stochastic
process with a spectral density having a power exponent $0<\alpha\le
2$. Among the many fascinating questions related
to it \footnote{a large collection of references is available online:\\ \href{\oofweb}{\texttt{\oofweb}}}
an apparently simple problem is how to build
random number generators (RNG) with such long-range correlations.
The classical method (see \eg \cite{ripley}) is to generate a vector of
white noise (as produced by a standard RNG) with an appropriate 
correlation matrix. However, one needs to take the ``square root'' of
this matrix (as with a Cholesky decomposition \cite{NR}), a method
rapidly limited by the size of the requested sample. 
One must then turn on to more specialized techniques, which can be
categorized into:
\begin{itemize}
\item
\textit{Digital signal processing}: one generate a vector of white noise and
fast-convolve it in Fourier space with the required \oofa shape (see
\eg \cite{fft}). One may use also a wavelet decomposition (\cite{wornell}) which
is quite natural due to the scaling properties of the process (see
next part). With these methods samples of the order of $o(10^8)$ can
be efficiently obtained. For samples with longer correlations, one is
limited by the memory of the computer.
\item \textit{Fractal techniques}: in non-linear science, \oofa noise
  is referred to as ``fractional brownian noise''.
The peculiar shape of the spectrum  (with a power exponent) implies that the
process is "scale-invariant" or more precisely "self-similar" (see \eg
\cite{fractals}), meaning 
that the stochastic process $X(t)$ is statistically similar (in the
sense of having the same statistical moments in the infinite limit) to
$\tfrac{X(r t)}{r^H}$ for any dilatation $r$, where $H$ is the Hurst
exponent related to the \oofa slope by $\alpha=2H+1$. Some very
efficient method to generate this process \cite{fournier} is 
to shoot Gaussian numbers for both ends of an interval, take the
average at the mean and add a new shot with a re-scaled amplitude,
and repeat the process on each subintervals. This however requires
knowing \textit{a-priori } the limits of your sample and is also
limited by the computer memory for very large samples.
\end{itemize}

In some cases, these methods are insufficient, as in some modern
astrophysics experiments. For instance, this work has been perform in the framework of the simulation of the
Planck ESA mission \cite{planck}
which is a satellite experiment, planned to
be launched in 2008, whose goal is to measure 
the Cosmic Microwave Background anisotropies with unprecedent
accuracy. It is composed of two instruments: High Frequency (HFI) and
Low Frequency (LFI), the former using low temperature bolometers (0.1K)
and the latter radiometers. At the level of precision required ($\simeq
10^{-5}K$) the instruments are sensitive to \ooff noise mostly from the 
cryogenic parts (HFI) and $\oofa$ with  $\alpha\simeq 1.7$ from the
low-noise electronics (LFI).  Given the sampling rate of the
instruments ($200$\hz) and its long duration (two years), very long
streams of noise need to be generated to prepare the analyzes . The
techniques described previously fail.

In the following we will show how to produce rapidly such long streams of
\oofa noise for an arbitrarily low frequency cutoff, without requiring
massive memory load.

In the \ooff case (section \ref{sec:HFI}) we will adapt some
classical numerical filtering technique (Auto-Regressive Moving
Average, see Appendix) that allows to build an \textit{optimal}
generator (\ie only limited by the underlying standard white one).
For $\alpha<2$  (section \ref{sec:LFI}) the problem becomes much more
intricate, because we are moving off the linear theory
and the previous method cannot be generalized.
We will then adapt a method from electronics, to
approximate the problem to a very good precision by a set of
\ooff generators and use the previous case.

We will then give hints on how to build generators with $\alpha>2$
(section \ref{sec:above}) using methods based on logistic maps.

The Appendix recalls some properties of the $z$ transform which will
be our main tool.

The algorithms described hereafter are now full part of the Planck
simulation programs \cite{reinecke} and
streams of 1 year data of noise ($\simeq 6.3~ 10^9$) are generated in
about 20 minutes on a single standard 2 GHz processor.


\section{ \ooff noise}
\label{sec:HFI}

\subsection{Why pure \ooff  noise is random walk.}

Let us start with the pure \ooff noise case. 

The idea is to start with a standard Gaussian generator and
numerically filter the shots  (called $x_i$, realizations of an $x(t)$
process) to obtain the proper power spectrum which is
simply \footnote{Through the article we will use the
  electronic notation for $j^2=-1$ and work in the frequency domain
  \w. In addition we will not mention the normalization factor $\sigma$ of the input
  Gaussian generator that can be recovered straightforwardly by
  multiplying the output by $\dfrac{\sigma}{\sqrt{\fsamp}}$.}:
\begin{equation}
  \label{eq:princip}
  S_y(\w)=|H(j\w)|^2 S_x(\w)
\end{equation}

Since $x$ is white $S_x=1$ , so all we need is to design a filter, here of the form:
\begin{equation}
  |H(j\w)|^2 = \dfrac{1}{\w^2} 
\end{equation}
therefore: 
\begin{equation}
  H(j\w) = \dfrac{1}{j\w} 
\end{equation}
This is simply a first order integrator and has an equivalent discrete
$z$ transform of the form (see Appendix):
\begin{equation}
  H(\zinv)=\dfrac{1}{1-\zinv}
\end{equation}
Therefore the filtered signal is:
\begin{equation}
\label{eq:rw}
  Y(\zinv)= H(\zinv) X(\zinv)=\dfrac{X(\zinv)}{1-\zinv}
\end{equation}
which according to the summation property of the $z$ transform
(Appendix) is
obtained simply from:
\begin{equation}
  y_k=\sum_{i=0}^k x_i
\end{equation}
This equation represents an MA (moving average) filter; the system is
built by "remembering" at each step its previous value (random walk) and has
therefore only one state variable. It has an "infinite memory".

It is slightly more convenient to write this filter in the following
way:
\begin{equation}
    (1-\zinv) Y(\zinv)=X(\zinv)
\end{equation}
which accordingly to the offset properties of the $z$ transform is
obtained in the time domain through:
\begin{equation}
  y_k-y_{k-1} = x_k
\end{equation}

which represents a one state AR (Auto Regressive) filter.

\subsection{Whitening below some minimum frequency \fmin}

While infinitely long streams may be theoretically interesting
(does pure \oofa noise has an intrinsic low frequency cutoff? Is 
it a stationary process?...) we are concerned here with designing
a generator for real life experiments
which have \textit{finite} durations. Therefore it is assumed that there exists
\textit{some} frequency $\fmin$ below which the noise becomes white
(which can be as long as the time-scale of the experiment).

To get a white spectrum below some frequency $\wz=2\pi \fmin$, 
we are seeking for a transfer function of the type:

\begin{eqnarray}
\label{eq:fmin}
  |H(j\w)|^2 &=& \dfrac{\wz^2}{\w^2+\wz^2}\\
  H(j\w) &= &\dfrac{\wz}{j\w+\wz}
\end{eqnarray}
which represents a first order low-pass filter.

Its inverse Fourier transform is:
\begin{equation}
  h(t)=\wz e^{-\wz t} u(t)
\end{equation}
($u(t)$ being the unit step), 
which sampled at a rate $1/T$ has the $z$ transform:
\begin{equation}
  H(\zinv)=\sum_i h(iT) z^{-i} = \dfrac{\wz}{1-e^{-\wz T}\zinv}
\end{equation}

Therefore:
\begin{equation}
  Y(\zinv) (1-e^{-\wz T}\zinv) = \wz X(\zinv)
\end{equation}

whose inverse transform is:
\begin{equation}
  y_k = \wz~ x_k + e^{-\wz T} y_{k-1} 
\end{equation}

For each shot $x_k$ one just needs to keep the previous filtered value
$y_{k-1}$ to compute the new one $y_k$.

\subsection{Whitening above some knee frequency \fk}
\label{sec:oof2w}

In most experimental cases, noise becomes white also above  some
"knee" frequency \fk and we incorporate it in the following.

For a filter with a white spectral behavior both below \fmin and above
\fk, we symmetrize Eq.\eqref{eq:fmin}:
\begin{eqnarray}
\label{eq:fkmin}
  |H(j\w)|^2 &=& \dfrac{\w^2+\wk^2}{\w^2+\wz^2} \\
  H(j\w) &= &\dfrac{j\w+\wk}{j\w+\wz}
\end{eqnarray}
In order to treat the numerator and denominator symmetrically, we use
the bilinear transform (see Appendix):
\begin{equation}
  w=j\dfrac{1-z}{1+z}
\end{equation}
relating it to frequency via:
\begin{equation}
  \w\thickapprox 2 w \fsamp
\end{equation}
to get the $z$ transfer function in the form:
\begin{equation}
  H(\zinv)=\dfrac{a_0+a_1\zinv}{1-b_1\zinv}
\end{equation}
with:
\begin{eqnarray}
\label{eq:coeff}
  a_0&=&\dfrac{1+r_1}{1+r_0} \nn \\
  a_1&=&-\dfrac{1-r_1}{1+r_0} \\
  b_1&=&\dfrac{1-r_0}{1+r_0} \nn 
\end{eqnarray}
where the reduced frequencies are $r_1=\pi\fk/\fsamp$ and
$r_0=\pi \fmin/\fsamp$. 

It is now straightforward to build the numerical filter from:
\begin{equation}
  Y(\zinv)(1-b_1\zinv)=(a_0+a_1\zinv)X(\zinv)
\end{equation}
which for any $k$ reads:
\begin{equation}
  y_k=a_0 x_k +a_1 x_{k-1} + b_1 y_{k-1}
\end{equation}

This is the equation for a simple ARMA filter: for any white shot
$x_k$ one just needs to keep the previous one $x_{k-1}$ and the
previous filtered value $y_{k-1}$, to compute the new one.

\begin{figure}[htbp]
  \centering
 \resizebox{.9\textwidth}{!}{\includegraphics{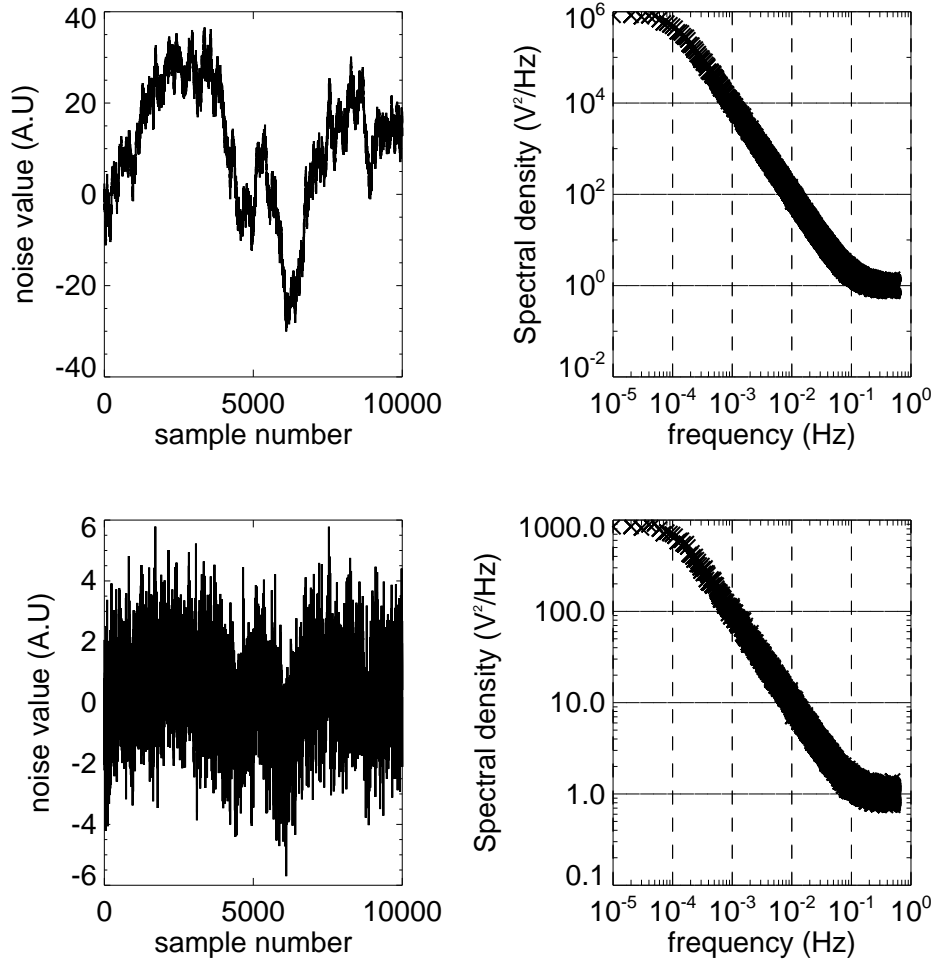}}
 \caption{\label{fig:oofres}  Results obtained with the generators.
   Upper plots: random walk ($\alpha=2$). Lower plots: $1/f$ noise
   ($\alpha=1$). $10^7$ samples are shot in each case. 
   A slice of the signal in real space is shown on the left, 
   and the power spectral
   density  on the right (estimated using a classical Welsh
   periodogram method \cite{NR}). The following bounds were used for whitening:
   $\fmin=10^{-4} \hz,\fk=10^{-1} \hz$. The logarithmic scale allows to
   check by eye the validity of the slope.}
  \end{figure}

This simple filter has been implemented in \texttt{C++} and run for $10^7$
samples. Results are depicted on the upper plots of Figure \ref{fig:oofres} 
and show that it behaves correctly.

This method is \textit{optimal},
in the sense that CPU-time is just limited by the speed of your
white number generator. It is insensitive to whatever
your maximal correlation length (\fmin) is.  And obviously it does not
require any memory load.

\section{ \oofa noise}
\label{sec:LFI}

\subsection{Failure of ARMA models}
\label{sec:failure}

ARMA models were so successful in the \ooff noise case that we are 
tempted to generalize them to the \oofa case. This is however far
from trivial, mainly because we are moving off linear theories for
which the $z$ transform has been tailored.
Let us see what happens in more details.

For a pure \oofa noise case,we are seeking for a transfer function of
the form:
\begin{eqnarray}
\label{eq:oofa}
  |H(j\w)|^2 &=& \dfrac{1}{\w^\alpha} \\
  H(j\w) &= &\dfrac{1}{(j\w)^{\alpha/2}}
\end{eqnarray}
Its generalized \cite{campbell} inverse Fourier transform is:
\begin{equation}
\label{eq:htoofa}
  h(t)=\dfrac{t^{-(1-\alpha/2)}}{\Gamma(\alpha/2)}
\end{equation}
and its associated one-sided $z$ transform is:
\begin{equation}
  H(\zinv)=\sum_{i \ge 0} h(iT)
  z^{-i}=\dfrac{T^{-(1-\alpha/2)}}{\Gamma(\alpha/2)}\sum_i
  i^{-(1-\alpha/2)} z^{-i}
\end{equation}
Unfortunately the sum that appears on the r.h.s cannot be factorized,
so that the MA filter $Y(\zinv)=H(\zinv) X(\zinv)$ leads
to an explicit convolution:
\begin{equation}
  y_k\propto \sum_{i=0}^{k} i^{-(1-\alpha/2)} x_{k-i}
\end{equation}

In direct space, one can therefore produce $1/f$  noise by generating
$\tfrac{1}{\sqrt{t}}$ type transients at random initial times \cite{radeka,ambrozy}.

There are however two problems with this expression:
\begin{enumerate}
\item the first term $i=0$ is infinite
\item we must keep \textit{all} the past terms ($x_{k-i}$)which is
  rapidly inefficient even with a fast convolution method.
\end{enumerate}

The first problem is the reflect of the high frequency divergence of
\oofa noise \cite{radeka}. Kasdin
\cite{kasdin} proposes an elegant way to circumvent it, by using a $z$
transfer function of the form : 
\begin{equation}
  H(\zinv)=\dfrac{1}{(1-\zinv)^{\alpha/2}}
\end{equation}
(compare to \eqref{eq:rw}). It allows us to build an AR filter:
$Y(\zinv)(1-\zinv)^{-\alpha/2}=X(\zinv)$ by the power expansion  
$ (1-\zinv)^{-\alpha/2}= \sum_k \xi_k z^{-k}$:
\begin{equation}
 \sum_{i=0}^{k} \xi_k y_{k-i}=x_k
\end{equation}
The $\xi_k$ tend rapidly to $k^{-(1+\alpha/2)}$ and the first term
$\xi_0=1$ is now well defined.

This however does not solve the long range dependency which just
reflects the fact that a \oofa ($\alpha<2$)
spectrum has \textit{an infinite number of state variables and that there
exits no (linear) difference equations to represent it}.

One may think that this bad behavior is to due to a missing low
frequency cutoff (\wz). We can modify the transfer function \eqref{eq:oofa} to
see what happens:
\begin{eqnarray}
  |H(j\w)|^2&=&\dfrac{\wz^\alpha}{(\w^2+\wz^2)^{\alpha/2}}\\
  H(j\w)&=&\dfrac{\wz^{\alpha/2}}{(j\w+\wz)^{\alpha/2}}
\end{eqnarray}
then using Eq. 3.382-7 \cite{gr} , the inverse Fourier transform is:
\begin{equation}
  h(t)=\dfrac{\wz^{\alpha/2}}{\Gamma(\alpha/2)} t^{-(1-\alpha/2)}
  e^{-\wz t}
\end{equation}
By comparing to the no-cutoff case Eq.\eqref{eq:htoofa}, we see that
the effect of introducing \wz is to attenuate the long
range correlation by the exponential factor: $\xi_k \ra \xi_k e^{-\wz k T}$.
Still, one needs to keep a few times $1/(\wz T)$ past samples which is
too inefficient for a low frequency cutoff.

\subsection{Recursive \ooff filtering}
\label{sec:recoofa}

We then turn on to an approximate method to produce efficiently long
range correlated samples. An old well-known way in electronics 
\cite{bernamont} is to
approximate the $1/f$ spectrum by a sum of \ooff,
\ie relaxation processes, equally spaced on a logarithmic frequency
grid, which is called the "infinite RC model" \cite{RC}.  Quite
surprisingly little filters are needed to produce an \oofa
($0<\alpha<2$) spectrum at the percent level. 
This method is used for instance in pink audio noise generation
\cite{audio}, in commercial libraries \cite{comm}, 
and even implemented on a DSP \cite{dsp}.

We work out here a slight variant of the RC model 
due to Keshner \cite{keshner} which allows to incorporate whitening both below \fmin and above \fk,
by filtering one single white shot. 

\begin{figure}[htbp]
  \centering
    \resizebox{.5\textwidth}{!}{\includegraphics{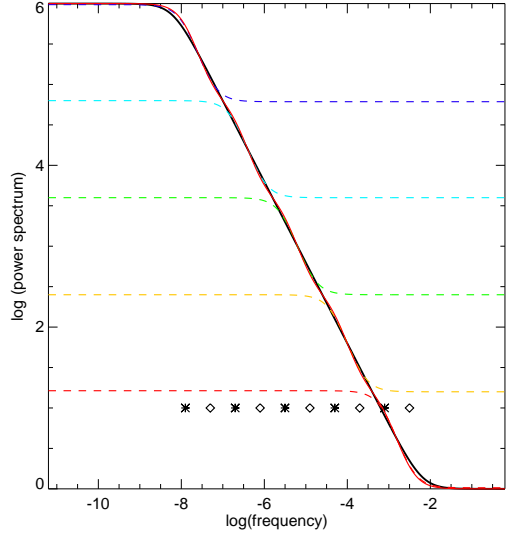}}
  \caption{A band-limited $1/f$ spectrum (full line) can be
    approximated by a set of \ooff spectra properly located (dashed
    lines) . The crosses and diamonds represent respectively the
    position of the poles and zeros in Laplace space. Here we use only
    one filter per
    decade; the approximation is already good.}
  \label{fig:keshner}
\end{figure}

We are seeking for a transfer function of the type:
\begin{equation}
\label{eq:fkamin}
  |H(j\w)|^2=\Bigl | \dfrac{\w^2+\wk^2}{\w^2+\wz^2} \Bigr |  ^{\alpha/2}
\end{equation}
The approximate transfer function is of the type:
\begin{equation}
  |H(j\w)|^2 \simeq \prod_{i=0}^{N_f-1} \dfrac{\w^2+z_i^2}{\w^2+p_i^2}
\end{equation}
where $N_f$ is the number of filters used, and  $p_i,z_i$ are the
poles and zeros locations.

Since we already designed these \ooff filters in section \ref{sec:oof2w}, the
only remaining point is to get their poles and zeros positions. 

From Figure \ref{fig:keshner} it is clear that the poles must be located regularly on a logarithmic grid\footnote{we use a base 10 logarithm.} :
\begin{eqnarray}
  \Delta p & = & \dfrac{\log \wk- \log \wz}{N_f} \\
  \log p_{i+1}& = &\log p_i + \Delta p
\end{eqnarray}
in order to have a symmetric errors, the zeros must be placed at \cite{saletti}:
\begin{equation}
  \log z_i =\log p_i +\dfrac{\alpha}{2}\Delta p
\end{equation}
and for continuity, from geometrical considerations, 
the first pole must be put at:
\begin{equation}
 \log p_0=\log \wz +\dfrac{1}{2}(1-\dfrac{\alpha}{2})\Delta p
\end{equation}

How many \ooff filters do we need? 
While in principle, as few as one
per decade is enough to approximate the slope to better than 5\%
\cite{keshner}, since we also wish here to reproduce neatly the \fmin
and \fk shoulders, we used 1.5 filters per decade, which leads to an
approximation at the percent level for any $0<\alpha<2$ value 
throughout the spectrum \footnote{The worst case being 
the $1/f$ spectrum.}.

We implemented this algorithm, using the previous code 
for \ooff noise: a Gaussian
white number is shot and passed recursively through the set of \ooff
filters: the coefficients of Eq \eqref{eq:coeff} are
automatically recomputed by the class for a given pole/zero 
and the output of each filter is the input to the next one.

We illustrate on the lower plots of Figure \ref{fig:oofres} results obtained by
generating $10^7$ samples of an $1/f$ noise.
It required using only six \ooff filters and 
the increase in CPU-time with respect to a pure white Gaussian shot, is
only of 25\%: the limitation is mainly due to the white time shot, not the
filtering stages.

The strong feature of this method is that the number of filters used
increases only logarithmically down to \fmin. The drawback is that the
generator has to be "warned up" to avoid the transient response, on a timescale of the order of the slowest RC filter (approximately
$1/\fmin$).

\section{$\alpha>2 ?$}
\label{sec:above}

While rarely encountered in experimental data,
the curious reader may wonder how to generate an \oofa
noise with a slope $\alpha>2$.  From geometric arguments, the methods
presented before fails. An interesting approach is to use the work on
logistic maps.

Consider the straightforward (non-linear) recurrence \cite{manneville}:
\begin{equation}
  x_{k+1}=x_k+x_k^2 ~(\mathtt{mod}~ 1)
\end{equation}
where $0<x_0<1$.

This is a pure $1/f$ noise generator (actually $1/f (\log f)^2 $)!


By changing the power 
exponent, group theory computations show \cite{ben} that any factor
$\alpha>2$ can be reached.  By adding a "shift from tangency"
$\epsilon$ in the recurrence: 
\begin{equation}
x_{k+1}=(1+\epsilon) x_k+ (1-\epsilon) x_k^2 ~(\mathtt{mod}~ 1)
\end{equation}
one introduces a low frequency cutoff \fmin.
Whitening above \fk can be obtained by shooting a properly weighted
extra white shot.

This chaotic
process is however far from Gaussian (it is an "intermittent" one,
representing long phases of low values, followed by sudden "bursts" of
high ones ) but, from the Central Limit Theorem, Gaussianity can be
recovered by summing up several such independent generators, a method 
particularly efficient on parallel computers, but beyond the scope of
this article.

For completeness, note an interesting connection between this approach
and a more empirical one consisting in re-scaling some Brownian motion (but
in which case only $0<\alpha<2$ range can be reached) \cite{gingl}.

\section*{Conclusion}

We have worked out the formalism of filtering white noise 
efficiently to produce "infinite" streams of band-limited \oofa
noise. 
\begin{itemize}
\item For \ooff noise the method is optimal, \ie is equivalent to
generating standard Gaussian white noise.
\item For $0<\alpha<2$ we use an excellent approximation which only
requires about one and half\ooff filters/decade, allowing the very fast
generation of \oofa noise over an arbitrarily large frequency range.
\end{itemize}
The corresponding (C++) software can be downloaded from:\\
\href{\randweb}{\texttt{\randweb}} \\

We have focussed mainly on this method because it is lightweight and
very efficient for very long time streams. 
For more general applications, before choosing a dedicated algorithm
ask yourself the following questions:
\begin{enumerate}
\item Do I want to generate a Gaussian noise?
\item What $\alpha$ range do I need?
\item Do I have a low frequency cutoff \fmin below which noise is
  white? Note that the upper \fk
  frequency can always be added by shooting an extra white
  noise (properly weighted).
\item On how many decades do I need the logarithmic slope?
\item What is my hardware support (mainly memory load)?
\end{enumerate}
Table \ref{tab:guide} then gives you some hints about which algorithm
you can use.

\begin{table}[htbp]
  \centering
  \begin{tabular}{|c|c|c|c|c|c|}
    \hline
    Method & Gaussian? & $\alpha$ range & \fmin & CPU/load & ref.\\
    \hline
    FFT/FWT & yes & any & yes & moderate& \eg \cite{fft}/\cite{wornell} \\ 
    Random pulses & yes & $0<\alpha \le 2$ & yes & high& \cite{radeka}, \cite{ambrozy} \\
    ARMA filtering & yes & $0<\alpha \le 2$ & yes & high& \cite{kasdin}, section \ref{sec:failure} \\
    Sum of \ooff filters & yes & $0<\alpha \le 2$ & yes & low & \cite{comm},\cite{audio},\cite{dsp},section \ref{sec:recoofa}\\
    Random midpoint displacement & yes & $1\le \alpha \le 3$ & no &
    low & \cite{fournier}, \cite{fractals}\\
    Brownian scaled motion & no & $0<\alpha \le 2$ & no &low &  \cite{gingl}\\
    Logistic maps & no & any & yes & low & \cite{manneville},\cite{ben} \\
    \hline
  \end{tabular}
  \caption{\label{tab:guide}
  Summary of today's existing methods to generate discrete \oofa noise. The
  column "\fmin"
  refers to the (possible) existence of a frequency limit in the
  method, beyond which noise
  is white. "CPU/load" is indicative (it should be considered as relative to
the different methods) and only significant for
very long correlation lengths.}
  \end{table}

\newpage
\appendix
\section*{Appendix}

In this appendix we recall a few results of the theory of automatism
that are relevant to this analysis. The reader is referred to any book
on linear theories (as \cite{shwartz}) for
more details.

The $z$ transform is a tool similar to the Fourier transform, but for
the discrete domain. To a causal series of values \serie{x_{i=0,1,..}}
we associate its $z$ transform by:
\begin{equation}
  X(\zinv)=x_0+x_1 \zinv +x_2 z^{-2}+...
\end{equation}
where $z$ is a complex number.

The following properties can then be demonstrated:
\begin{eqnarray}
\label{eq:proptz}
\text{linearity}& \lambda_1\serie{x_i} + \lambda_2 \serie{y_i} \tz
\lambda_1 X(\zinv) + \lambda_2 Y(\zinv) \nn \\ 
\text{offset} & \serie{x_{i-k}}\tz z^{-k}X(\zinv) \\
\text{summation} & \serie{\sum_{k=1}^i x_k} \tz  
\dfrac{X(\zinv)}{1-\zinv} \nn \\
\text{convolution} & \serie{\sum_{k=0}^i x_k y_{i-k}} \tz X(\zinv) Y(z^{-1}) \nn
\end{eqnarray}

A time signal $x(t)$ sampled at a period $T$ is described as:
\begin{equation}
  x^\ast(t)=\sum x(iT) \delta(t-iT)
\end{equation}

and we can associate to the samples $x_i=x(iT)$ a $z$ transform. Some usual
transforms are:
\begin{eqnarray}
  \delta(t) & \tz & 1 \\
  u(t) &\tz & \sum_{i=0}^\infty z^{-i}=\dfrac{1}{1-\zinv}\\
  e^{\wz t} u(t) & \tz  & \sum_{i=0}^\infty(e^{\wz T}z)^{-i}=\dfrac{1}{1-e^{\wz T} \zinv}
\end{eqnarray}

An interesting property is related to energy conservation. The
power of the \textit{sampled} signal is related to its $z$ transform:

\begin{equation}
  \lvert X^*(j\w)\rvert ^2 ={X(\zinv) X(z)}_{|z=e^{j\w T}}
\end{equation}

The link between $\lvert H^\ast(j\w) \rvert$ and $\lvert H(j\w)
\rvert$ being not convenient, it is sometimes interesting to use the
$w$ bilinear (or Tustin) transform which is defined as
\begin{equation}
  w=j \dfrac{1-z}{1+z}
\end{equation}

It can be shown that it is related to the frequency of the signal \w
by
\begin{equation}
  w=\tan (\w T/2)
\end{equation}

In the limit of fast sampling $\w T \ra 0$ (which is most often the
case since we work below Nyquist frequency) the $w$ transform can 
be identified with \w, giving a convenient way to determine the $z$
transform coefficients directly from the signal transfer function.

Finally, from the properties \eqref{eq:proptz}, a rational polynomial $z$
transfer function of the form
\begin{equation}
  H(\zinv)=\dfrac{Y(\zinv)}{X(\zinv)}=\dfrac{a_0 +a_1 \zinv+...}{1+b_1\zinv+...}
\end{equation}

leads to the following filtering in the discrete time domain:
\begin{equation}
  y_k + b_1 y_{k-1}+...=a_0 x_k + a_1 x_{k-1}+...
\end{equation}

when only $a_0$ is different from 0, it is an called Auto regressive (\textbf{AR})
filter (also know as Infinite Impulse Response). If all the $b_i$'s are
null, this is a Moving Average (\textbf{MA}) one (also know as Finite Impulse Response). In the general case this is called an \textbf{ARMA} filter.

\label{lastpage}

\end{document}